\documentclass[Royal,times,sagev]{sagej}

\usepackage{amsmath, amsthm, amscd, amsfonts, amssymb, graphicx, xcolor, float, algorithm, algorithmic, subcaption, booktabs}
\usepackage[bookmarksnumbered, colorlinks, plainpages]{hyperref}
\usepackage{multirow}
\usepackage{caption} 

\floatstyle{plain}
\newfloat{mybox}{tbp}{lob}
\floatname{mybox}{Box}
\restylefloat{mybox}

\theoremstyle{definition}

\theoremstyle{remark}

\title{Calculating the Expected Value of Sample Information accounting for missing data}

\author{
Luca Benetti$^{1}$,
Gianluca Baio$^{2}$,
Anna Heath$^{3,4}$\\[1em]
{\small $^{1}$Human Technopole, Milan, Italy}\\
{\small $^{2}$Department of Statistical Science, University College London, London, UK}\\
{\small $^{3}$The Hospital for Sick Children, Toronto, Canada}\\
{\small $^{4}$University of Toronto, Toronto, Canada}\\
}

\begin{document}


\begin{abstract}
\textbf{Background}: The Expected Value of Sample Information (EVSI) is a powerful instrument to determine the value of additional evidence to inform a health economic model. However, EVSI has been applied only to idealized data collection mechanisms, without considering the potential challenges that may arise in realistic studies, thereby reducing its potential applications. In this paper, we define a methodology to calculate EVSI when the additional evidence we aim to collect exhibits missing data; a very common challenge in real-world studies.\\
\textbf{Methods}: First, we define how to simulate individual-level data and how to induce missingness inside the simulated data. We will reproduce Missing Completely At Random (MCAR), Missing At Random (MAR), and Missing Not At Random (MNAR) missing data mechanisms. Then, we will apply the multiple imputation method to adjust for the bias related to the missing data. Finally, we use the imputed data to compute the EVSI using nonparametric regression methods.\\
\textbf{Results}: We will apply the novel methodology to two different health economic models and compare the EVSI computed on idealized data (without missingness) with the EVSI computed with different types of missing data. We will show that the EVSI decreases when the additional evidence suffers from missingness, and therefore, we define a method to efficiently compute the sample size we need to collect to recover the idealized EVSI (on data without missingness). We find out that the number of additional samples needed to correct for missing data exceeds that coming from standard approaches in our applications.\\
\textbf{Conclusions}: With this methodology, we compute EVSI when the additional data we aim to
collect are affected by non-trivial forms of missingness, modeling both the MAR and
MNAR missingness mechanisms, and extending EVSI calculation to more realistic scenarios. The fact that the necessary sample size to recover the idealized EVSI exceeds that of standard methods suggests that this methodology potentially represents a novel technique for sample size calculation in realistic studies.
\end{abstract}


\Highlights{
1. We propose a methodology to extend the calculation of EVSI to account for complex missing data mechanisms in additional data, allowing EVSI computation in realistic scenarios.\\
2. When data are simulated with MAR and MNAR missingness mechanisms, EVSI reduces due to additional sources of uncertainty. \\
3. We propose a computationally efficient method to calculate the necessary sample size to reach the original level of EVSI (without missingness), finding out that the number of additional samples needed to correct for missing data exceeds that coming from standard approaches.}


\maketitle

\section{Introduction}
The Expected Value of Sample Information (EVSI) \cite{ades2004expected} quantifies the economic value of collecting additional evidence to reduce uncertainty in an underlying health-economic model. In general, additional evidence will decrease uncertainty in the parameters that characterize the economic model, ultimately leading to updated knowledge about which treatment is optimal in terms of cost-effectiveness. By computing EVSI, we can quantify the economic value of this uncertainty reduction.

EVSI has been shown to be a powerful tool for determining whether further evidence is worthwhile in many applications \cite{kunst2019value}\cite{reeves2019three}. It prevents resources from being wasted on (i) interventions that only appear to be effective due to limited information and (ii) research into interventions that have already been evaluated with sufficient precision. It also helps researchers to design studies by clarifying (i) which data should be collected to reduce the uncertainty in the parameters that are driving decision uncertainty (research prioritization) and (ii) how much additional data should be collected to maximize the value of information while minimizing the associated costs.

EVSI has generally been applied to idealized data collection exercises \cite{heath2022simulating, heath2024value}. The most common study design is a randomized controlled trial (RCT) that computes the efficacy of the different interventions, meaning that EVSI measures the value of reducing uncertainty for the treatment effect estimate \cite{tuffaha2016value} \cite{pei2023value} \cite{doble2017economic}. Specific methods have also been developed to compute EVSI for different outcome types, such as survival outcomes \cite{vervaart2022efficient}, and study designs, such as cluster randomised trials \cite{welton2014expected}. However, these methods do not account for challenges in the data collection process, such as the existence of missing data.

Understanding how to perform EVSI while accounting for missing data can be beneficial, as it enables us to utilize these powerful methods in a more realistic context. Moreover, utilizing EVSI incorporating missing data would be helpful to design experiments when we know that the proposed data collection strategy will suffer from missingness. Methods to simulate data with missingness for EVSI computation have already been developed \cite{heath2022simulating}; however, the existing approach is only applicable when the data are assumed to be missing completely at random (MCAR); meaning that, in a Bayesian perspective, the prior knowledge about the underlying missing generating mechanisms is that it is MCAR. Notice that this is a highly unrealistic assumption for real-world studies, where the missing data generating mechanism is usually more complex. Thus, we will develop a novel methodology to compute EVSI while accounting for more realistic missing data.

To compute EVSI with realistic missing data, we start by generating individual-level data from a proposed RCT. We then introduce different assumptions for the missing data. Then, we analyze the incomplete data using various techniques to account for missingness \cite{little2019statistical} \cite{rubin2018multiple} \cite{seaman2013review}. Finally, we compute EVSI on the ``adjusted'' data. In this work, we focus on using multiple imputation to recover complete data before computing EVSI. We can then compare EVSI with and without accounting for missing data and evaluate how many additional patients are required to adjust for the assumed missingness. 

This paper proceeds as follows: first, we define EVSI,  clarify the taxonomy of missing data, and specify how missing data can be generated and addressed through multiple imputation. We then introduce our general framework to compute EVSI with missing data. Third, we apply this methodology in two different health-economic models, a simple Normal-Normal conjugate model for the data-generating process and a realistic Markov model \cite{heath2024value}. Finally, we discuss how to effectively utilize this methodology and more accurately compute EVSI in a realistic context.

\section{Methods}
\label{section 2}

\subsection{Expected Value of Sample Information}
EVSI computes the value of collecting data to reduce uncertainty in the parameters of an underlying health economic model. As such, consider a health economic model that is aiming to compare a set of $D$ different interventions, $d=1,\dots,D$. Each intervention, $d$, is valued using the net monetary benefit \cite{stinnett1998net}, $\text{NB}_d$. The net benefit is calculated using a set of parameters $\boldsymbol{\theta}$, $\text{NB}_d(\boldsymbol{\theta})$. These parameters are typically estimated from data, meaning that they are subject to uncertainty. We define $p(\boldsymbol{\theta})$ as the joint distribution that represents the uncertainty associated with the set of parameters. In a standard Bayesian decision framework, to choose the best decision, one has to maximize $\mathbb{E}(\text{NB}_d(\boldsymbol{\theta}))$. This is performed by first sampling $N$ values from $p(\boldsymbol{\theta})$ and then computing $\mathbb{E}(\text{NB}_d(\boldsymbol{\theta}))$ via Monte Carlo (MC) simulation \cite{metropolis1949monte}. The process of sampling $N$ values from $p(\boldsymbol{\theta})$ is known as either \textit{probabilistic sensitivity analysis} (PSA) or \textit{probabilistic analysis} (PA)\cite{heath2024value} .

Within this context, EVSI can be then defined as follows:
\[
\text{EVSI} = \mathbb{E}_{\boldsymbol{Y}}\left[ \max_{d} \mathbb{E}_{\boldsymbol{\theta} | \boldsymbol{Y}}\left( \text{NB}_d(\boldsymbol{\theta})\right) \right] - \max_{d} \mathbb{E}_{\boldsymbol{\theta}}\left( \text{NB}_d(\boldsymbol{\theta})\right),
\]
where $\boldsymbol{Y}$ represents the data we plan to collect to inform our decision model \cite{ades2004expected}. The outer expectation is taken with respect to the prior predictive distribution defined through $p(\boldsymbol{Y}\mid\boldsymbol{\theta})$ and $p(\boldsymbol{\theta})$ \cite{heath2022simulating}. 

In practice, EVSI must be computed using efficient approximation methods, such as the regression-based method, moment matching or importance sampling \cite{heath2017review} \cite{heath2018efficient}. In this work, we compute EVSI using the regression-based method \cite{strong2015estimating}, which uses non-parametric regression to compute the EVSI. 

For the non-parametric approach to EVSI calculation, one first has to simulate dataset $\boldsymbol{y}$ from the prior predictive distribution $\boldsymbol{y} \sim p(\boldsymbol{Y})$ by first sampling parameters $\boldsymbol{\theta}^* \sim p(\boldsymbol{\theta})$ and then $\boldsymbol{y} \sim p(\boldsymbol{Y}|\boldsymbol{\theta}^*)$. Once data are simulated, multiple nonparametric regression equations (one for each possible treatment) are fitted to compute the preposterior mean $\mathbb{E}_{\boldsymbol{\theta} | \boldsymbol{Y}}\left( \text{NB}_d(\boldsymbol{\theta})\right)$:

\[
\text{NB}_d(\boldsymbol{\theta}) =  g_d(T(\boldsymbol{y})) + \epsilon,
\]
where $\boldsymbol{\theta}$ is a sample of the joint distribution of the input parameters obtained by performing a probabilistic analysis (PA), and $T$ is some low-dimensional summary statistic of the data that summarizes the information in the data. Ideally, $T$ should include all the information in the original data; for this reason, it is usually suggested to choose sufficient statistics of the parameters we assume are informed by the data \cite{li2024estimating}. Other reliable options, when sufficient statistics are not obtainable, are typically sample estimators or maximum Likelihood estimators (MLE) of the parameter(s) we expect to inform collecting new data \cite{li2024estimating}. Under this approach $\mathbb{E}_{\boldsymbol{\theta} | \boldsymbol{Y}}\left( \text{NB}_d(\boldsymbol{\theta})\right) \approx \hat{g}_d\left(T(\boldsymbol{y})\right)$ and EVSI can be computed as:
\[
\text{EVSI} = \frac{1}{K}\sum_{i = 1}^K \max_d \hat{g_d}(T(\boldsymbol{y}^{(i)})) - \max_d  \frac{1}{K}\sum_{i = 1}^K \hat{g_d}(T(\boldsymbol{y}^{(i)})).  
\]

Current implementations of EVSI use simplified data collection schemes, meaning that $\boldsymbol{Y}$ does not resemble real data. This work aims to introduce a methodology that enables us to compute the value of collecting $\boldsymbol{Y}$ when the data exhibit realistic patterns of missing data. To achieve this, we must first introduce methods to simulate realistic RCT data before discussing different types of missing data that can occur and how the complete RCT data can be adjusted for each type of missingness.

\subsection{Simulating data with missingness}

\subsubsection{Modeling the complete randomised controlled trial}
\label{simulate_full_data}

To generate data with realistic missingness, we need to specify a model to generate a complete dataset with individual-level synthetic RCT data \cite{heath2022simulating}. In many cases, the parameters governing the sampling distribution of the data may not be available in the health economic model and must be derived from the literature or expert knowledge. If the study team has access to data from a preliminary RCT, this information can also be leveraged to generate future study data.

In general, we assume that our proposed RCT will collect information from $n$ individuals. This information will include $S$ covariates $\{\boldsymbol{X} = X_1, X_2,.., X_S$\}, the exposure variable $Z$ and the outcome variable $Y$. To model RCT data, we factorize the joint distribution of the covariates and the outcome variable into a product of sequential conditional distributions. In particular, for $X_1$, we first define the marginal distribution as $X_1 \sim F_1(\boldsymbol{\Theta}_0)$. We then proceed by defining the conditional distribution of $X_2 | X_1$ as $X_2 | X_1 \sim F_2(\boldsymbol{\Theta}_1, X_1),$ and recursively defining the other distributions for the remaining covariates, $F_s$, for $s = 1, 2, \dots, S$. 

The exposure variable $Z$ can be randomly assigned, e.g., $Z_j \overset{iid}\sim  \text{Bernoulli}(q)$, for $j=1,..,n$, where $q$ is the expected proportion of treated individuals. Equivalently, the first $nq$ individuals can be assigned a $1$ (indicating treated), while the remaining $n(1-q)$ can be assigned $0$. Finally, for each individual, we define the distribution of the outcome given the covariates and the exposure variable as $Y | \boldsymbol{X}, Z \sim F(\boldsymbol{\Theta}(\boldsymbol{X}, Z))$, where $\Theta$, the relevant parameters of $F$, depend on both $\boldsymbol{X}$ and $Z$. Moreover, when computing EVSI, the distribution of the outcome must also depend at least one of the model parameters underlying the health economic model $\boldsymbol{\theta}$. 

In this way we have defined the distribution for the RCT data $(\boldsymbol{Y}_{\text{RCT}}, \boldsymbol{X}, \boldsymbol{Z})$; notice that here, for $\boldsymbol{Y}_{\text{RCT}}$ we refer to the distribution of the $n$ individuals' outcomes, for $\boldsymbol{X}$, the distribution of the $n$ individuals' covariates, i.e., a matrix of dimension $n\times S$, and for $\boldsymbol{Z}$ the distribution of the $n$ individuals' exposure variables. The simulated RCT data will be denoted as $(\boldsymbol{y}_{\text{RCT}}, \boldsymbol{x}, \boldsymbol{z})$. Now, we must determine the type of missingness we wish to introduce. There are different missing data mechanisms \cite{rubin1976inference}\cite{little2019statistical}, which each require an alternative method to reproduce in simulated data.  

\subsubsection{Missing completely at random}\label{simulate_MCAR}
The simplest missing data mechanism is missing completely at random (MCAR), which assumes that the probability that an individual has missing data is independent of the covariates $\boldsymbol{X}$ and on the outcome $Y$. Given this definition, MCAR data can be simulated by sampling a missingness indicator $\pi_j$ for $j=1,..., n$ individuals, \[\pi_j \sim \text{Bernoulli}(p_0),\] with $p_0$ the expected percentage of missingness. Finally, to create a dataset with missingness, simply set $Y_j = \text{NA}$ for those individuals with $\pi_j = 1$ \cite{heath2022simulating}. Unfortunately, this simple missing data mechanism is also unrealistic as, in real-world scenarios, there is nearly always a relationship between an individual's characteristics and their probability of missing data. 

\subsubsection{Missing at random}

The missing at random (MAR) mechanism \cite{rubin:1987} is more complex and realistic and assumes that the missingness depends on the covariates $\boldsymbol{X}$ but not on the outcome $Y$. This means that the individual probability of missingness $p_j$, for $j=1,\dots, n$, can be defined as follows:

\[
    p_j = G_{\text{MAR}}(\beta_{00} + G_{\text{MAR}}^{-1}(p_0) + g_{\text{MAR}}(\boldsymbol{\beta},\boldsymbol{X}_j)),
\]

\noindent where $p_0$ is the expected percentage of missingness,  $\boldsymbol{\beta}$ are coefficients of the relationship between the covariates and the missingness, $g_{\text{MAR}}$ is the functional form of the relationship between the covariates and the missingness, and $G_{\text{MAR}}$ is the link function of the assumed binary model. Note that when $g_{\text{MAR}} = \sum_{s=1}^S \beta_sX_s$ is linear and $G_{\text{MAR}}(z,\mu, \tau) = \frac{1}{1 + e^{-\frac{z - \mu}{\tau}}}$ is the CDF of a standard logistic distribution, we recover the standard logit model\cite{berkson1944application}. In contrast, if $G_{\text{MAR}}$ is the CDF of a standard Normal distribution, we recover the probit model\cite{bliss1934method}. Finally, the parameter $\beta_{00}$ is an intercept that must be estimated with numerical methods so that the following equation is satisfied:
    \[
    \frac{1}{n}\sum_{j=1}^nG_{\text{MAR}}(\beta_{00} + G_{\text{MAR}}^{-1}(p_0) + g_{\text{MAR}}(\boldsymbol{\beta},\boldsymbol{x}_j)) - p_0 = 0.
    \]
This ensures that the expected percentage of missing outcomes in the generated data is $p_0$. Note that a systematic review estimated that the average proportion of missing individuals in applied studies is approximately $0.2$, with some cases where the proportion exceeded $0.6$\cite{fiero2016statistical}. Finally, as for MCAR, the final dataset is created by sampling the missingness indicator $\pi_j$ from a Bernoulli distribution $\pi_j \sim  \text{Bernoulli}(p_j)$, and setting $y_j = \text{NA}$ for those individuals having $\pi_j = 1$.

\subsubsection{Missing not at random}

Finally, the missing not at random (MNAR)\cite{molenberghs2007missing} \cite{schafer2002missing} mechanisum assumes that the missingness depends on both the covariates $\boldsymbol{X}$ and the outcome $Y$. To produce an MNAR model, we define $p_j$ as:
    \[
    p_j = G_{\text{MNAR}}(\gamma_{00} + G_{\text{MNAR}}^{-1}(p_0) + g_{\text{MNAR}}(\boldsymbol{\gamma},\boldsymbol{X}_j,Y_j))
    \]
where, $g_{\text{MNAR}}$ represents how the probability of missingness depends on the covariates $\boldsymbol{X}_j$, the parameters $\boldsymbol{\gamma}$ and the outcome $Y_j$, and $G_{\text{MNAR}}$ is the appropriate link function. Similarly, $\gamma_{00}$ is estimated with numerical methods so that
    \[
    \frac{1}{n}\sum_{i=1}^nG_{\text{MNAR}}(\gamma_{00} + G_{\text{MNAR}}^{-1}(p_0) + g_{\text{MNAR}}(\boldsymbol{\gamma},\boldsymbol{x}_j,y_j)) - p_0 = 0,
    \]
to ensure that the expected percentage of missing outcomes in the generated data is $p_0$. Again, the  missingness indicator can be simulated from a Bernoulli distribution  and setting $y_j=$ NA for individuals with $\pi_j = 1$. As EVSI with MCAR data has already been considered, we will focus solely on calculating EVSI for MAR and MNAR data generating mechanisms. 

\subsubsection{Analyzing for missing data}

To analyze data with missing observations, we typically have several choices \cite{gabrio2017handling}. The simplest method is to discard all individuals with missing outcomes and analyze the available data only, known as a complete case analysis. This method yields inaccurate and/or biased estimates of the quantities of interest, except under MCAR, due to both analyzing a smaller amount of data (which increases variance in the estimate) and overlooking the fact that missingness can be associated with specific relevant characteristics of individuals. If the data are MAR, then the dependence between the outcome and covariates can be estimated using multiple imputation methods \cite{rubin:1987}\cite{rubin1987multiple} to provide unbiased estimates. In particular, this method involves imputing missing observations multiple times using different techniques to estimate the conditional predictive distribution of the missing values given the observed characteristics. After generating different complete imputed datasets, they are combined, using the Rubin rule or other techniques, to obtain relevant estimates of the missing values; in this way, we can use the imputed data in the EVSI computation process to compute an unbiased estimate of $T(\boldsymbol{Y})$ \cite{rubin1987multiple}. Finally, if the data are assumed to follow the MNAR mechanism, then accurate, unbiased analyses can only be achieved by modeling the missingness explicitly, which jointly models the missingness indicator variable and the outcome.

\subsection{Calculating EVSI accounting for missing data}
\label{EVSI with missing data}

\subsubsection{Simulating the covariates of interest}
\label{chapter3_covariate}

To compute EVSI, we must first simulate data with realistic missingness. The first step is to generate covariates for each individual $\boldsymbol{x}_j$, $j =1, \dots, n$, from the conditional distributions $F_s, s = 1, \dots, S$. We also simulate the exposure variable $z_j$ for $j = 1, \dots, n$. These covariates will be fixed throughout the EVSI calculation, meaning that our simulated datasets from the RCT will have different outcomes for each PA parameter $\boldsymbol{\theta}^{(i)}$, $i = 1, \dots, K$ but the same covariate profile. As the covariates are the same, EVSI is computed conditional on the exact simulated covariate profile. As such, we require an additional step to compute the unconditional EVSI, which marginalizes out the dependence on the covariates. In practice, this means that EVSI must be computed $M$ times before being averaged to compute the final EVSI estimate. 

\subsubsection{Simulating the patient-level outcomes}
Conditional on the simulated values for $\boldsymbol{x}$ and $z$, the trial outcomes must be simulated. Recall that the data generating model for the outcomes is conditional on the PA simulations $\boldsymbol{\theta}^{(i)}$, $i = 1, \dots, K$ \cite{heath2022simulating}. As such, the outcomes will be generated $K$ times, one dataset for each PA simulation. Moreover, we want the data-generating model to be also dependent on the covariates. To impose this dependence, we define $\tilde{\boldsymbol{\theta}}^{(i)}$ as a function of $\boldsymbol{\theta}^{(i)}$ and $\boldsymbol{X}$, and we consider the data-generating model for the outcomes conditional directly on $\tilde{\boldsymbol{\theta}}^{(i)}$. Explicitly,

    \[
    \tilde{\boldsymbol{\theta}}^{(i)} = f(\boldsymbol{\theta}^{(i)},\boldsymbol{X}),
    \]
where $f$ is a function representing the dependence structure between individual outcomes and covariates. For instance we may have that $\tilde{\boldsymbol{\theta}}^{(i)}$ is a linear combination of $\boldsymbol{\theta}^{(i)}$ and $\boldsymbol{X}$:

    \[
    \tilde{\boldsymbol{\theta}}^{(i)} = \boldsymbol{\theta}^{(i)} + \sum_{s=1}^S \beta_s(X_s - \overline{X})
    \]

We will denote the complete simulated data for each PA simulation $i = 1, \dots, K$ as $(\boldsymbol{y}^{(i)}_{RCT}, \boldsymbol{x},\boldsymbol{z})$, noting again that $\boldsymbol{x}$ and $\boldsymbol{z}$ are fixed across all PA simulations.

\subsubsection{Generating data with missingness}

Having simulated $(\boldsymbol{y}^{(i)}_{RCT}, \boldsymbol{x},\boldsymbol{z})$, we must then determine the appropriate missingness mechanism and simulate which data are missing. This is denoted $(\boldsymbol{y}^{(i)}_{\text{miss}}, \boldsymbol{x}, \boldsymbol{z})$, for $i=1, \dots, K$. In this case, $\boldsymbol{y}^{(i)}_{\text{miss}}$ is a subset of $\boldsymbol{y}^{(i)}_{\text{RCT}}$ such that $p_0$ values are missing. Once $\boldsymbol{y}^{(i)}_{\text{miss}}$ are sampled, we have successfully simulated data from a realistic RCT.

To calculate EVSI using the regression based method, we require a summary measure for the data. When accounting for missing data, we must summarise $\boldsymbol{y}^{(i)}_{\text{miss}}$ such that it provides an unbiased estimate of $T(\boldsymbol{Y}^{(i)}_\text{RCT})$. In general, the same summary function of the data with missingness, $T(\boldsymbol{y}^{(i)}_{\text{miss}})$, will not provide an unbiased estimate of $T(\boldsymbol{Y}^{(i)}_\text{RCT})$. As such, we must perform multiple imputation, or another relevant method, to compute $\tilde{T}(\boldsymbol{y}^{(i)}_{\text{miss}})$, which is an unbiased estimate of  $T(\boldsymbol{Y}^{(i)}_\text{RCT})$. 

Finally, having simulated the data, EVSI can be computed using the non-parametric regression method. Specifically, we fit a regression between $\text{NB}(\boldsymbol{\theta}^{(i)})$, as the dependent variable, and $\tilde{T}(\boldsymbol{y}^{(i)}_{\text{miss}})$, as the independent variable. The fitted values are then extracted from this regression $\hat{g_d}(\tilde{T}(\boldsymbol{y}^{(i)}_{\text{miss}}))$ before being used to estimate of 

\begin{equation}
\label{eq1}
\text{EVSI} | \boldsymbol{x} = \frac{1}{K}\sum_{i = 1}^K \max_d \hat{g_d}(\tilde{T}(\boldsymbol{y}^{(i)}_{\text{miss}})) - \max_d  \frac{1}{K}\sum_{i = 1}^K \hat{g_d}(\tilde{T}(\boldsymbol{y}^{(i)}_{\text{miss}})).  
\end{equation}

Finally, we integrate out the dependence on $\boldsymbol{x}$ using MC computation and obtain EVSI:
\[
\text{EVSI} = \mathbb{E}_{\boldsymbol{X}}(\text{EVSI} | \boldsymbol{X}) \approx \frac{1}{M}\sum_{m=1}^M(\text{EVSI} | \boldsymbol{x}_{\text{m}}),
\]
where $\boldsymbol{x}_{\text{miss}}$ is the simulated covariate profile for the $m-$th dataset, $m= 1, \dots, M$. This process is summarised in the Box \ref{box:evsi}.

\begin{mybox}[htbp]
\centering
\fbox{%
  \begin{minipage}{0.95\textwidth}
    \begin{algorithmic}[1]
    
    \FOR{$m = 1, \dots, M$}
        \STATE Generate covariates $\boldsymbol{X}_{\text{m}}$ and the exposure variables $\boldsymbol{Z}_{\text{m}}$ for $n$ individuals.
        
        \FOR{$i = 1, \dots, K$}
            \STATE Sample parameter set $\boldsymbol{\theta}^{(i)}$.
            \STATE Compute the net benefit function for each decision, $\text{NB}(\boldsymbol{\theta}^{(i)})$.
            \FOR{$j = 1, \dots, n$}
                \STATE Create individual parameter values $\tilde{\theta}^{(i)}_j$
                \STATE Sample individual outcomes:
                \[
                 y_{j}^{(i)} \sim F(\tilde{\theta}^{(i)}_j).
                \]
                and obtain $(\boldsymbol{y}^{(i)}_{RCT}, \boldsymbol{x}_{\text{m}},\boldsymbol{z}_{\text{m}})$
            \ENDFOR
            \STATE Based on the assumed missing mechanism, remove data to generate missing data.
            \STATE Use appropriate method to compute $\tilde{T}(\boldsymbol{y}^{(i)}_{\text{miss}})$, an unbiased estimator of $T(\boldsymbol{Y}^{(i)}_{\text{RCT}})$.
        \ENDFOR
        
        \STATE Fit a nonparametric regression model between
        $\{\text{NB}(\boldsymbol{\theta}^{(i)})\}_{i=1}^K$ and $\{\tilde{T}(\boldsymbol{Y}_{\text{miss}}^{(i)})\}_{i=1}^K$ 
        \STATE Extract fitted values and compute $\text{EVSI}(\boldsymbol{X}_{\text{m}})$ as in Equation \eqref{eq1}.
    \ENDFOR
    
    \STATE Integrate out dependence on $\boldsymbol{X}_{\text{m}}$ to obtain the final estimate of the EVSI.
    
    \end{algorithmic}
  \end{minipage}
}
\vspace{6pt}
\caption{Simulazione del processo EVSI}
\label{box:evsi}
\end{mybox}

\subsubsection{Adjusting sample size to account for missingness}
We will show that, as expected, EVSI computed on data with missing observations will be lower than EVSI computed with the complete dataset. This is partially due to the fact that the higher the sample size of simulated data, the higher the EVSI \cite{heath2019estimating}. In many study design scenarios, we are interested in adjusting the study sample size to account for missingness, i.e., determining the study sample size that would provide the same value of information as a sample with no missing data. As such, we now propose a method to compute the required sample size adjustment for the assumed missingness mechanism.  

For the MAR missingness mechanism, this adjustment for missingness can be achieved by increasing the sample size of the data we plan to collect. The straightforward way to compute the required sample size to provide the same value of information as an RCT with all the observed data is to sequentially increase the sample size until the original EVSI value is obtained. However, this can be computationally expensive. Therefore, we propose an efficient method to obtain the required sample size increase. This method works by identifying a sample size such that the variance for $\tilde{T}(\boldsymbol{Y}_{\text{miss}})$ matches the variance of the original $T(\boldsymbol{Y}_{\text{RCT}})$. This requires us to increase the sample size of $(\boldsymbol{y}_{\text{miss}}, \boldsymbol{y}, \boldsymbol{z})$ and compute the summary statistic through multiple imputation but avoids the need to perform the non-parameteric regression and compute EVSI. This method works by ensuring that the both the mean (since $\tilde{T}$ is unbiased) and variance of the summary statistics are the same, which provides a similar EVSI estimate \cite{heath2018efficient}.

\section{Application}

In this section, we describe two models used to compute EVSI with missing data. The first one assumes a Normal-Normal model for simple computation while the second uses a realistic underlying health-economic model that combines a Markov state transition model and a decision tree model \cite{heath2024value}.

\subsection{Normal-Normal model}
\label{EVSI in presence of missing data}

For the probabilistic analysis, we assume we have information on the mean outcome in the treatment and control groups. Both distributions are normal distribution with means $\theta_1 = 40$ and $\theta_0 = 20$, and variances $\sigma_1^2= 8$ and $\sigma_0^2 =6$ for the treatment and control groups, respectively. Thus, formally, the priors for the mean outcome, conditional on $Z$, are:
 \[
    \theta_{Z=0}^{(i)} \sim N(\theta_0,\sigma_0^2) \quad i=1,..,K
    \]
    \[  
    \theta_{Z=1}^{(i)} \sim N(\theta_1,\sigma_1^2) \quad i=1,..,K.
    \]    
Our proposed study aims to collect data from $n = 100$ participants with $n_0$ individuals receiving the control and $n_1$ individuals receiving the treatment. For simplicity, we set $n_0 = n_1$. The study outcomes are normally distributed with variance $\sigma^2_{\boldsymbol{Y}}$ and individual means, based on their covariate profiles. The net benefit functions for this model are given by:
\begin{equation}
\begin{aligned}
\text{NB}_0(\theta) = k\theta_0\\
\text{NB}_1(\theta) = k\theta_1 - c
\end{aligned}
\end{equation}

We assume each individual has two covariates, age and number of comorbidities $(age, com)$. For each individual, $j = 1,\dots, n$, $age$ is generated from a normal distribution, $age_j \overset{\mathrm{iid}}{\sim} \mathcal{N}(\mu_{age},\sigma^2_{age})$ while $com_j \overset{\mathrm{iid}}{\sim} Bin(n_{com},p_{com}) \quad j=1,...,n$ with $\mu_{age} = 57$, $\sigma_{age} = 8$, $n_{com} = 10$ and $p_{com} = 0.5$. 

For each simulated individual $j$, we compute their mean outcome, given the covariates, $\tilde{\theta}_j$ as follows:
\[
\tilde{\theta}^{(i)}_j = \theta_{Z_j}^{(i)} + \alpha_1(age_j-\overline{age}) + \alpha_2(com_j-\overline{com}) \quad j=1,...,n
\]
where $\theta^{(i)}_{Z_j}$ is the $i$-th PA simulation for group $Z_j$, with the coefficient $\alpha_1 = 3$ and $\alpha_2 = 5$ and 
\[
Y_j^{(i)} {\sim} N(\tilde{\theta}_j^{(i)}, \sigma_Y^2).
\]
For this example, we take $K = 5000$ and obtain $(\boldsymbol{y}_{\text{RCT}}^{(i)},\boldsymbol{x}, \boldsymbol{z})$ for $i=1,..,K$. 

In this example, we set $p_0 = 0.5$ as the expected percentage of missingness in the data and generate data for both the MAR and MNAR mechanisms. For each simulated individual $j$, we compute $p_j$ for the MAR models as \[
p_j = \Phi(\beta_{00} + \Phi^{-1}(p_0) + \beta_1(age_j-\overline{age}) + \beta_2(com_j-\overline{com})),\] with $\beta_1 = 0.06,\beta_2 = 0.1$ and $\beta_{00}$ estimated with numerical methods. Similarly, for the  MNAR model we define $p_j$ as \[p_j = \Phi(\gamma_{00} + \Phi^{-1}(p_0) + \beta_1(age_j-\overline{age}) + \beta_2(com_j-\overline{com}) + \beta_3(y_j - \overline{y})),\] with $\beta_1 = 0.06, \beta_2 = 0.1, \beta_3 = 0.2$ and $\gamma_{00}$ estimated with numerical methods.
    
Once we have the probability of a missing outcome $p_j$ for each individual $j$, we then sample the missingness indicator from a Bernoulli model and set $y_j =$ NA for $\pi_j=1$. Finally, to estimate the treatment effect from the missing data, we perform multiple imputation, using a generalised linear model and 20 imputations. For MAR, this provides an unbiased estimate of outcome mean in each treatment that is $\tilde{T}(\boldsymbol{y}_{\text{miss}})$. The mean outcome is a sufficient statistic of the parameters $\theta^{(i)}$, which means that EVSI can be accurately estimated \cite{li2024estimating}.

Finally, we fit a Generalized additive model (GAM) to obtain an estimate of $\text{EVSI} | \boldsymbol{X}_{\text{m}}$ before integrating out the dependence on $\boldsymbol{X}$ to obtain the final estimate of the EVSI.

\subsection{Chemotherapy treatment model}
\label{markov_missing}

The second application uses a more realistic model \cite{heath2024value}, which combines two common structures in health economic modeling: a decision tree model and a Markov state transition model. In this example, the decision maker must select from $D=2$ different chemotherapy treatments. The therapies do not differ in terms of efficacy, but in terms of their costs and side effects profiles, with the more expensive therapy inducing fewer side effects. The model estimates the proportion of individuals who experience $s_d$ after receiving standard care, $d=0$, or novel treatment, $d=1$. The Markov model represents the possible trajectories of the health state and their related costs that individuals with side effects might incur.

The future study will collect information on $s_0$ and $s_1$. In this application, we consider a sample size of $n = 1000$. We simulate the same two independent covariates for each individual $j$, using the same distributions as for the normal-normal model with $n_0=n_1=500$. The information on the probability of experiencing side effects with the two chemotherapy options is assumed to come from primary data (for $d = 0$) and the literature (odds ratio for the treatment effect $\rho$). All other parameters and probabilistic analysis distributions are listed in \cite{heath2024value}.
    
For each individual $j$, we compute $\tilde{s}^{(i)}_j$, the probability of experiencing side effects given the patient's $j$ characteristics) as follows:
    \[
    \tilde{s}^{(i)}_{Z_j,j} = \Phi( \Phi^{-1}(s_{Z_j}^{(i)}) + \alpha_1(age_j-\overline{age}) + \alpha_2(com_j-\overline{com}) + \alpha_{0,Z_j})
    \]
    where $s_0^{(i)}$ is the $i$-th sample of PA simulation (for $d=0$), $\alpha_1 = 0.05$, $\alpha_2 = 0.3$, $\Phi$ is the CDF of a standard Gaussian and $\alpha_{0,Z_j}$ is calculated to ensure that the underlying probability of side effects equals $s_0$ or $s_1$, depending on the exposure variable for the individual $j$. The outcome for individual $j$ is simulated as:
    \[
    Y_j^{(i)} \overset{\mathrm{iid}}{\sim}  \text{Bernoulli}(\tilde{s}_j^{(i)})
    \]
    to obtain simulated RCTs $(\boldsymbol{y}_{RCT}^{(i)},\boldsymbol{
    x},\boldsymbol{z})$ for $i=1,..,K$.

We define the missingness mechanisms using the sample distributions as the Normal Normal model. The summary statistic $T$ will then be the odds ratio of side effects, which will be computed using multiple imputation for the missing data and directly from the RCT to compute EVSI without missingness. 

\subsection{Comparison}
\label{Comparison and second moment matching}

For each of the two examples, we will take $K=5,000$ as the number of PA simulations. We will also set $M = 10$ as the number of covariate profiles simulated and 20 multiply imputed datasets will be used to generate the unbiased statistic. We will present the following quantities:

\begin{enumerate}
    \item EVSI computed on the simulated RCT data with no generated missingness, known as the Original EVSI. To compute EVSI, we summarise $\boldsymbol{Y}_{\text{RCT}}$ using either the mean or the odds ratio.

    \item EVSI computed by assuming that $\boldsymbol{Y}_{\text{miss}}$ is the complete data. This is equivalent to performing a complete case analysis and will directly summarise $\boldsymbol{Y}_{\text{miss}}$ using either the mean or the odds ratio. This provides a biased summary statistic.
    
    \item EVSI when missingness is generated under the MAR assumption and multiple imputation is used to adjust for the missingness. Note that under the  MAR assumption, multiple imputation gives unbiased summary statistics.

    \item EVSI when missingness is generated under the MNAR assumption and multiple imputation is used to adjust for missingness. Note that, multiple imputation returns biased estimates under MNAR assumptions as the missingness also depends on partially observed variables.

\end{enumerate}

To present the EVSI results, we will plot the values across a range of willingness to pay parameters $k$. We will also summarise the differences between the EVSI values by computing the average difference between the Original EVSI and the other EVSI values, across the range of willingness-to-pay values $k$.

\section{Results}
\subsection{Normal-Normal model}

From top to bottom, Figure 1 displays the original EVSI (without missingness), the EVSI with MAR missing mechanism, the EVSI on complete data, and the EVSI with MNAR missing mechanism for the normal-normal model.

       \begin{figure}
        \centering
        \includegraphics[width=1\linewidth]{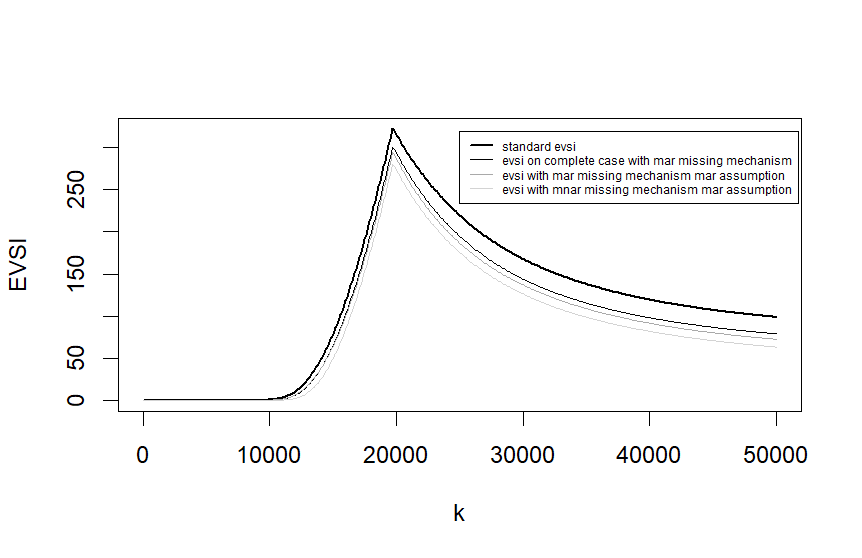}
        \caption{From top to bottom: EVSI calculated relying on RCT data $(\boldsymbol{Y}_{\text{RCT}}^{(i)},\boldsymbol{X}, \boldsymbol{Z})$ for $i=1,..,K$,  EVSI computed on multiple imputed data $(\boldsymbol{Y}_{\text{miss}}^{(i)}, \boldsymbol{X}, \boldsymbol{Z})$  under MAR assumption, EVSI on fully observed data (complete case), EVSI on  multiple imputed data $(\boldsymbol{Y}_{\text{miss}}^{(i)}, \boldsymbol{X}, \boldsymbol{Z})$  under MNAR assumption.} 
    \end{figure}

EVSI decreases when the data are affected by missingness. This is mainly due to the additional variability introduced by the missingness mechanism and the multiple imputation used to recreate the missing data. The first source of variability is related to the fact that before collecting additional data, we do not know which data will be missing, and we must account for this additional uncertainty in our model. The second is related to the additional variance added by the multiple imputation. 

We see that EVSI using the complete case analysis is lower than EVSI computed using multiple imputation. This means that multiple imputations partly compensates for the loss in precision that we incur when analysing the data with missingness. The final comparison from Figure 1 demonstrates that EVSI computed when the data are MNAR, is lower than the EVSI computed by performing a complete case analysis. However, the challenge here is that the MNAR missing mechanism results in lower precision and biased summary statistics. Thus, this behaviour must be further investigated to better understand the effects of an unbiased estimate of $T(\boldsymbol{Y}_{\text{RCT}})$ in the EVSI computation.

Figure 2(a) is used to determine the sample size required to provide the same value as the complete RCT with $n=100$ recruitment. The horizontal blue line represents the value of var($T(\boldsymbol{Y}_{RCT})$), while the dashed lines represent +/- 0.025 of the variance.

\begin{figure}[h]
    \centering
\begin{subfigure}{0.48\textwidth}
        \centering
        \includegraphics[width=\linewidth]{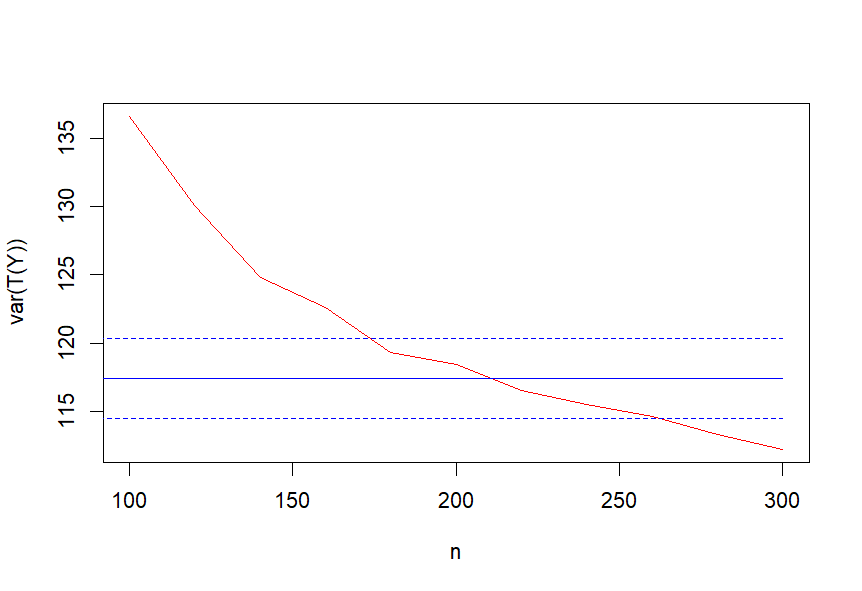}
        \caption{}
        \label{fig:a}
    \end{subfigure}
    \hfill
    \begin{subfigure}{0.48\textwidth}
        \centering
        \includegraphics[width=\linewidth]{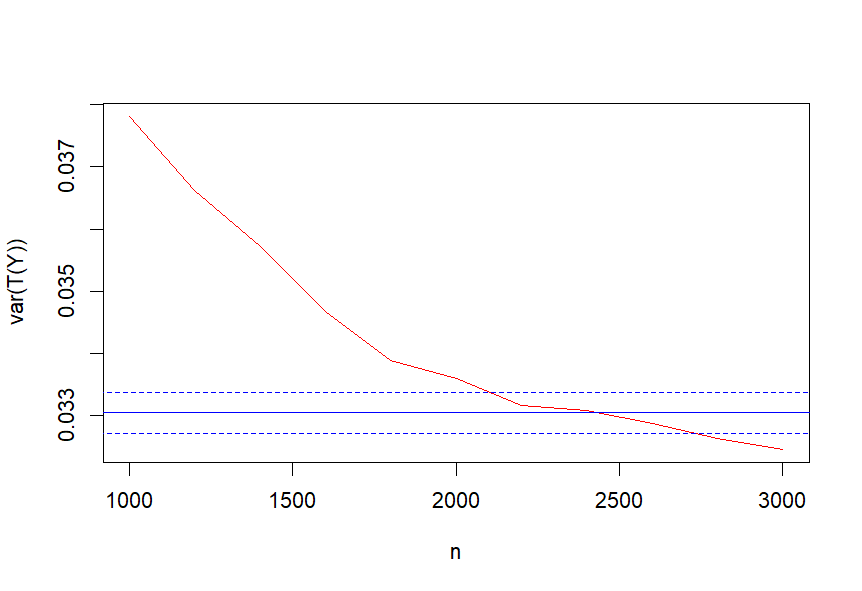}
        \caption{}
        \label{fig:b}
    \end{subfigure}
    
    \caption{Comparison between var($\tilde{T}(\boldsymbol{Y}_{\text{miss}})$) for increasing sample size $n$ and var($T(\boldsymbol{Y}_{\text{RCT}})$): (a) Normal-Normal model, (b) Chemotherapy treatment model.}
    \label{fig:comparison}
\end{figure}

\noindent From Figure 2(a), we can identify that $n_{adj} \approx 210$ results in a similar variance estimate between $(T(\boldsymbol{Y}_{\text{RCT}}))$ and ($\tilde{T}(\boldsymbol{Y}_{\text{miss}})$). Thus, the sample size must be adjusted from 100 to 210 to provide a similar value of information. To demonstrate that EVSI is accurately adjusted with this sample size, Table 1 represents the mean difference (absolute and relative to the Original EVSI estimate) between the Original EVSI and  (i)  EVSI assuming a MAR missing mechanism, (ii)  EVSI on complete data, (iii)  EVSI assuming MNAR missing mechanism, and (iv) EVSI assuming an MAR missing mechanism with adjusted sample size. The willingness to pay was varied at $k_0 = 50$ and $k_1 = 200$.

From Table 1, we see that the Original EVSI value is close to the EVSI with the adjusted sample size, while all others have a substantial relative mean different.  Note that the original sample size was $n=100$ with a percentage of missingness $p_0 = 0.5$, and the adjusted sample size was estimated as $n_{adj} \approx 210$. Typically, in studies affected by missingness, the adjustment researchers make follows this relation $n_{adj} \approx n/(1-p_0)$, meaning that using a standard approach to sample size calculation in this specific scenario, we would have collected data with a sample size of $200$, which slightly underestimates the value of the information.

\subsection{Chemotherapy treatment model}

Figure 3 displays the original EVSI (without missingness), the EVSI with MAR missing mechanism, the EVSI on complete data, and the EVSI with MNAR missing mechanism for the Chemotherapy treatment model. We observe similar patterns to the normal-normal model with the multiple imputation recovering some of the value compared to the complete case analysis but still losing value compared to the EVSI with $\boldsymbol{Y}_{\text{RCT}}$. The MNAR provides the lowest value, despite having the same level of overall missingness. 

\begin{figure}
        \centering
        \includegraphics[width=1\linewidth]{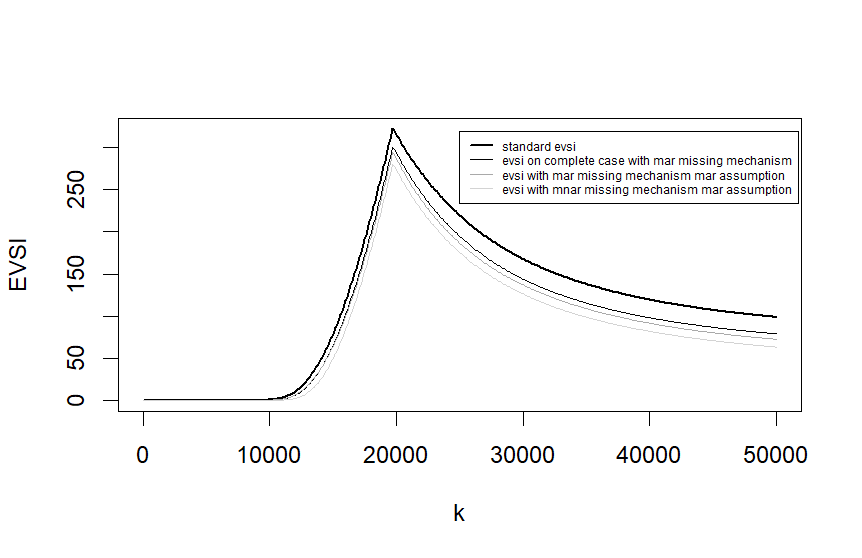}
        \caption{From top to bottom: EVSI calculated relying on RCT data $(\boldsymbol{Y}_{\text{RCT}}^{(i)},\boldsymbol{X}, \boldsymbol{Z})$ for $i=1,..,K$,  EVSI computed on multiple imputed data $(\boldsymbol{Y}_{\text{miss}}^{(i)}, \boldsymbol{X}, \boldsymbol{Z})$  under MAR assumption, EVSI on fully observed data (complete case), EVSI on  multiple imputed data $(\boldsymbol{Y}_{\text{miss}}^{(i)}, \boldsymbol{X}, \boldsymbol{Z})$  under MNAR assumption.}    \end{figure}

In Figure 2(b), we display the results to obtain the adjusted sample size. Based on these results, we set $n_{adj} \approx 2450$ and compute the EVSI with missing data and an MAR mechanism. Table 1 includes the absolute and relative mean differences between the EVSI values with $k_0 = 15000$ and $k_1 = 30000$.

\begin{table}[h]

\centering
\begin{tabular}{rrrrrr}
  \toprule
Example &Measure&$\Delta_{orig,mar}$ & $\Delta_{orig,comp}$ & $\Delta_{orig,mnar}$ & $\Delta_{orig,adj}$ \\ 
  \midrule
\multirow{2}{*}{Normal-Normal} & MD & 25.68 & 44.39 & 54.61 & 1.14 \\ 
&Relative MD & 0.13 & 0.23 & 0.29 & 0.01 \\  

  \midrule
\multirow{2}{*}{Chemotherapy}& MD & 22.17 & 28.75 & 41.25 & 4.47 \\ 
&Relative MD & 0.10 & 0.13 & 0.19 & 0.02 \\ 
  \bottomrule
\end{tabular}
\caption{Mean and relative differences between the original EVSI and the EVSI assuming a MAR missing mechanism, the EVSI on complete data, the EVSI assuming MNAR missing mechanism and the EVSI assuming a MAR missing mechanism computed on the adjusted sample size $n_{adj}$. First line Normal-Normal model, $k_0 = 50$ and $k_1 = 200$. Second line Chemotherapy treatment model, $k_0 = 15000$ and $k_1 = 30000$.}
\label{tab:comparison}
\end{table}
\noindent Similar to before, the Original EVSI has been achieved (within 2\%) by matching the variance of $\tilde{T}(\boldsymbol{Y}_{\text{miss}})$ with $T(\boldsymbol{Y}_{RCT})$. Note that the original sample size was $n=1000$ with $p_0 = 0.5$. However, the adjusted sample size is $n_{adj} \approx 2450$. The standard sample size adjustment would recommend a sample size of $2000$, which is substantially lower than the sample size required to recover the original EVSI. 

\section{Discussion}

In this paper, we introduced a method that incorporates more complex data collection mechanisms into the calculation of EVSI. We focused on defining a method to compute EVSI when researchers expect that the data will be affected by missingness. Critically, we focused on settings where the missingness is assumed to be informative and depend on covariates or even the outcome itself. This allows for the application of EVSI to more realistic applications.

When computing EVSI with data that account for missingness, EVSI is reduced through two different sources of uncertainty. Firstly, EVSI decreases as additional variance is introduced by the analysis method we employ to handle missing data, specifically multiple imputation. Secondly, there is the uncertainty in the missing mechanism, since we don't know in advance which observations will be missing. As such, we have seen that the standard sample size adjustment used in RCTs does ensure that the value of the data is sufficient when comparing to an RCT with no missing data. We have also developed a computationally efficient method to determine the number of individuals required to match the information from an RCT with no missingness.

Future research could explore the theoretical properties of EVSI with missing data to better define, identify, and quantify the additional sources of uncertainty. These properties would be required to determine a formal technique for quantifying the loss in EVSI due to the different types of missing generating mechanisms (MCAR, MAR, MNAR), or, in other words, the economic value of missing data. Another interesting avenue for future research would be to further explore the MNAR missing mechanism. Critically, multiple imputation is not sufficient to obtain an unbiased estimate of the statistic $T$. This makes it challenging to apply the nonparametric method to compute EVSI as the summary statistic introduces an additional layer of uncertainty.

Finally, alternative methods to multiple imputation could be tested to analyse the data with missing observations. In this way, we can evaluate which method is better at retrieving the initial level of EVSI, thereby defining another way to compare different methods used to deal with missingness.

\subsection{Conclusion}
We introduced a novel methodology to compute EVSI when the additional data we aim to collect are affected by non trivial forms of missingness. We modeled both the MAR and MNAR missing mechanisms, and we analyze the effects of multiple imputation method on EVSI, by imputing missing data and computing EVSI on complete data. We showed that the value of EVSI decreases because of mainly two specific layers of additional uncertainty introduced by (i) the multiple imputation estimates and (ii) the uncertainty in the missing mechanism; and proved that in different scenario the standard adjustment usually made for handling missingness in health economic contexts does not work. Finally, we constructed a computationally efficient method to calculate the necessary sample size to reach the original level of EVSI (without missingness). This research opens up EVSI to more realistic scenarios, extending its calculation to real-world data behaviors.

\renewcommand{\bibname}{References}
\bibliographystyle{plain}
\bibliography{bib1.bib}

\end{document}